# Strain-engineering the Schottky barrier and electrical transport on MoS$_2$

Ashby Phillip John, Arya Thenapparambil and Madhu Thalakulam[1]

School of Physics, Indian Institute of Science Education & Research Thiruvananthapuram, Kerala, India 695551

## Abstract

Strain provides an effective means to tune the electrical properties while retaining the native chemical composition of the material. Unlike three-dimensional solids, two-dimensional materials withstand higher levels of elastic strain making it easier to tune various electrical properties to suit the technology needs. In this work we explore the effect of uniaxial tensile-strain on the electrical transport properties of bi - and few-layered MoS$_2$, a promising 2D semiconductor. Raman shifts corresponding to the in-plane vibrational modes show a redshift with strain indicating a softening of the in-plane phonon modes. Photoluminescence measurements reveal a redshift in the direct and the indirect emission peaks signaling a reduction in the material bandgap. Transport measurements show a substantial enhancement in the electrical conductivity with a high piezoresistive gauge factor of ~ 321 superior to that for Silicon for our bi-layered device. The simulations conducted over the experimental findings reveal a substantial reduction of the Schottky barrier height at the electrical contacts in addition to the resistance of MoS$_2$. Our studies reveal that strain is an important and versatile ingredient to tune the electrical properties of 2D materials and also can be used to engineer high-efficiency electrical contacts for future device engineering.

---

[1] madhu@iisertvm.ac.in

**Introduction**

Strain in the crystal lattice, by altering the lattice constants, orbital interaction and coupling between the neighboring atoms provides a powerful means to engineer the physical properties of a solid. Carrier mobility enhancement in strained Si on Silicon–Germanium substrates offered a vital solution for complementary metal-oxide-semiconductor (CMOS) performance improvements[1,2]. An elastic strain of 1% is reported to induce a change of ~100 meV in the material bandgap[3]. In three-dimensional high-tensile materials such as stainless steel, only a small fraction ~ <1% of the applied strain results in useful elastic deformation of the lattice while the rest relaxes via plastic deformations such as defects dislocations[2,4]. Contrary to the bulk materials, lower dimensional systems are shown to withstand higher levels of elastic strain[5,6]. Owing to the two-dimensional nature the van der Waals (vW) materials are shown to host higher degrees of in-plane strain[7,8]. Graphene has been reported to withstand the highest amount of strain[7] while the transition metal di-chalcogenides (TMDC) following closely[8]. Among the TMDCs, $MoS_2$, a promising microelectronics material, boasts a surprisingly high Young's modulus and have shown to withstand much larger amounts of strain (~ 10%)[8,9] compared to many of the high tensile materials[2]. The possibility of engineering the bandgap, carrier mobility, carrier density, effective mass etc., by the application of strain can establish $MoS_2$ as a potential microelectronic and straintronic material.

Reports on strain-tuning physical properties on TMDCs have been focussed mostly on spectroscopic studies such as Raman scattering and photoluminescence (PL) studies[10–14]. A continuous reduction of the bandgap, effective mass and a transition from the direct to the indirect bandgap has been predicted in TMDCs[15–17]. It has been shown that the bandgap of $MoS_2$ can be continuously tuned by applying in-plane strain[11,13,18]. In addition, a direct to indirect bandgap transition and metal-insulator transition are also observed on $MoS_2$[11,13] and $MoTe_2$[19].

Potential for device applications of any phenomena require its manifestation in electrical transport experiments. Such reports, where the effect of strain on the electrical transport properties such as the conductivity, the band gap and the metal-semiconductor contact barrier heights on TMDCs are only a handful[18–25]. Most of the studies utilize an atomic force microscope (AFM) tip[18,19,21,22] to apply local strain, while transport studies on the global application of strain are only a few[23–26]. On $MoS_2$, the strain applied by the AFM tip propagates only up to a distance of ~ 500 nm[18] laterally while an accurate study of the bandgap, effective mass, the Schottky barrier height (SBH) at the contacts etc., requires the application of strain uniformly across the sample. Moreover, the tip-induced modifications to the transport properties also need to be addressed. Bending the substrate in a direction normal to its plane exploiting a single or a double ended beam geometries[23,24] and, stretching or compressing the substrate[25] are the most common methods used to apply uniform strain on TMDCs. Piezoelectric and piezoresistive effects[23,25] have been observed on $MoS_2$ by these methods.

MoS$_2$ forms Schottky junctions with commonly used contact materials making the realization of efficient electrical contacts challenging[27–30]. Though the SBH is predicted to depend on the work-function of the metal and the electron affinity of the semiconductor[31], the dominant Fermi-level pinning mechanism[30,32] makes the SBH reduction utilizing lower work-function metals ineffective[28,33]. It has been shown by density functional calculations that the in-plane strain can lower the SBH and the contact resistance mediated by bandgap reduction and improved hybridization of electronic orbitals of the $MoS_2$ and the contact metal[34]. A complete study requires addressing both the changes to the electronic properties such as the bandgap and the behavior of the electrical contacts against uniform strain, which forms the major focus of this report.

In this work the effect of uniaxial strain on the bandgap, electrical conductivity and the SBH at the electrical contacts of bi- and few-layered $MoS_2$ devices is explored. We utilize a three-point bending apparatus to exert the strain. The strain induced on to the sample is characterized using Raman spectroscopy. The modulation of the bandgap with strain is quantified using PL spectroscopy. Low-noise electrical transport measurements on the devices reveal a high piezoresistivity with a gauge factor (GF) larger than that for Silicon[35]. Our observations show both the conductivity of the material and also the SBH at the electrical contacts undergoes a substantial reduction with strain. Our studies reveal that strain is an important handle to tune the electrical properties of the material and also can be used to engineer high-efficiency electrical contacts for future device engineering.

**Methods**
Commercially procured bulk $MoS_2$ crystals (SPI supplies) are mechanically exfoliated with the help of an adhesive tape and subsequently transferred on to a 0.5 mm thick flexible substrate (polyimide, Sigma-Aldrich) using the PDMS assisted dry transfer technique[36]. A uniaxial strain is transferred onto the $MoS_2$ flake by bending the flexible substrate normal to its plane exploiting a micrometre-controlled three-point bending apparatus. A schematic drawing of the bending mechanism is shown in Fig. 1(a). From the thickness $2t$ and bending radius $\rho$ of the substrate, as shown in Fig. 1(b), we calculate the strain induced on the substrate and the sample, $\varepsilon = t/\rho$[11]. A plot of the strain vs the number of turns on the micrometre screw is shown in the Fig. 1(c). Raman and PL spectroscopy are performed using Horiba Xplora Plus spectrometer with a 532 nm excitation with 2400 lines/mm grating and 600 lines/mm respectively. Electrical contacts to the flakes are defined using photo-lithography followed by Cr (2nm)/Au (30 nm) metallization.

**Results & Discussion**

Tensile strain results in the softening of the lattice spring constants and in-turn the crystal phonon modes. Raman spectroscopy is an effective tool to study the phonon modes and the variation in the Raman shifts has been used to study and characterize strain in $MoS_2$[10,12,37]. Here, we utilize the Raman spectroscopy to identify and characterize the strain exerted on to the $MoS_2$ flake from the bending apparatus. Raman spectra representing the prominent vibrational modes, $E^1_{2g}$ and $A_{1g}$, for the mono- and the bi-layer $MoS_2$ flakes for various percentage of applied tensile strain are shown in Fig. 1(d) and (e) respectively. The inset to Fig. 1 (d) illustrates the vibrational modes; the $E^1_{2g}$ mode ~ 380 cm$^{-1}$ corresponds to the in-plane vibrational mode while the $A_{1g}$ mode ~ 400 cm$^{-1}$ corresponds to the out-of-plane vibrational mode. Raman spectra is also used to probe the number of layers in the $MoS_2$ flake in the few-layer (<4 layers) limit[11,38–40]. In this work the primary estimation of the thickness of $MoS_2$ flake is performed using optical contrast[13,41] and the Raman spectra prior to the application of strain on to the sample [red open-circles in Fig. 1 (d) and (e)] is used to confirm the number of layers. We obtain a shift of ~19.2 cm$^{-1}$ and ~22.03 cm$^{-1}$ for the unstrained mono- and bi-layer samples respectively; these values are consistent with the reported shifts for the mono- and the bi-layer $MoS_2$ samples on similar substrates[11,38,39], [See Supporting information S1]. The optical images of the corresponding $MoS_2$ flakes are shown in Supporting Information S2.

From Fig. 1 (d) and (e), we observe a redshift in the $E^1_{2g}$ peak corresponding to the in-plane vibrations of the Mo and S atoms indicating a softening of the mode. Fig. 1(f) and (g) show the variation in the Raman shifts for both the $E^1_{2g}$ and the $A_{1g}$ modes for the mono- and the bi-layer samples as a function of strain, extracted from the Fig. 1 (d) and (e) respectively. A linear fit to the shift gives a variation of ~ 2.62 ± 0.35 cm$^{-1}$/%-strain and 1.87 ± 0.1 cm$^{-1}$/%-strain for the $E^1_{2g}$ mode for the mono- and the bi-layer samples respectively. Strain transferred from the substrate is maximum for the adjacent layer and decreases for subsequent layers. As a result, the shift in the Raman mode with strain reduces with layer number[10].

The $A_{1g}$ peak shows negligible dependence on the strain. The $A_{1g}$ mode, corresponding to the out-of-plane vibrations of the Mo and S atoms, depends on the stiffness of the out of plane spring-constants, and is relatively unaffected for these ranges of in-plane strain values[38]. The observed dependence of the Raman modes with strain are consistent and comparable to those reported elsewhere[10,12,37] confirming that the strain estimated from the geometry of the bending apparatus is accurate with in the experimental uncertainties.

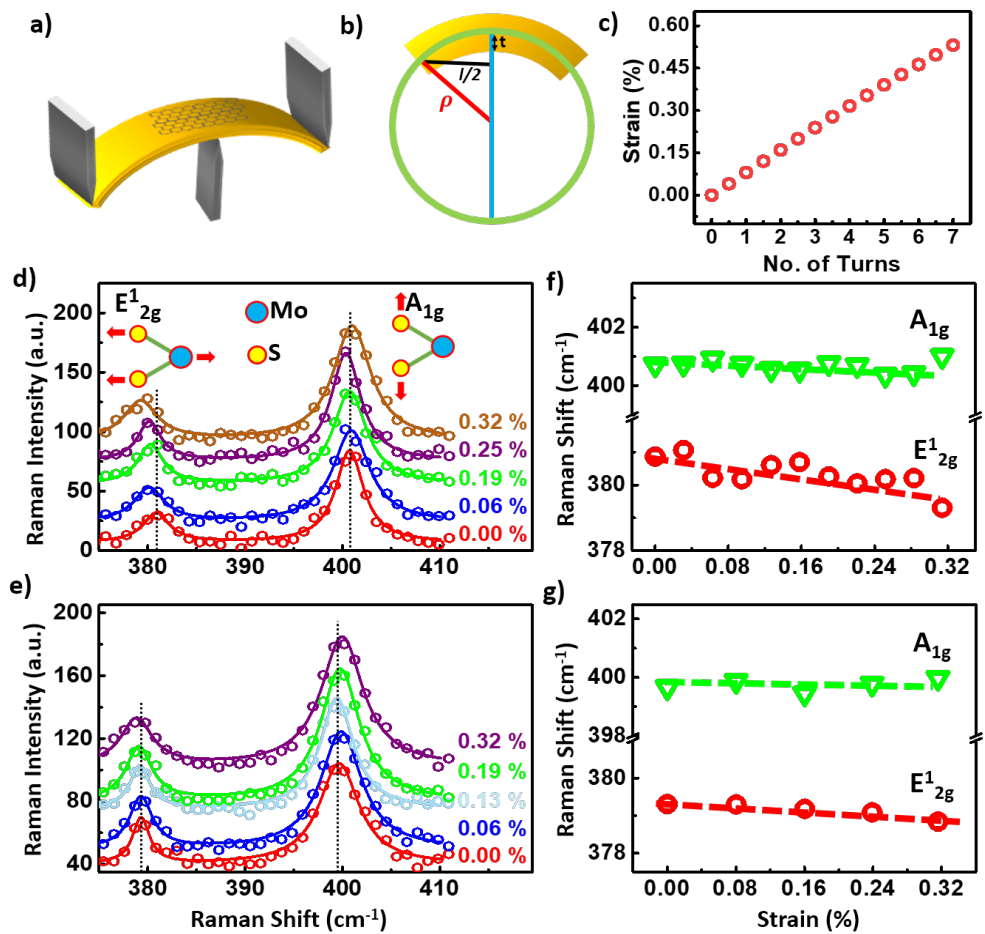

**Figure 1.** (a) Schematic representation of the bending apparatus. A flexible polyimide sheet of width 2 cm and thickness 0.5 mm is clamped between two knife edges with a distance of 4.5 cm between them. The centre of the substrate is deflected to apply the strain using a calibrated micrometre screw. (b) Geometrical representation of the strained substrate with thickness. The yellow arc represents the substrate with thickness $2t$, length $l$ and $\rho$ is the radius of curvature. The convex side of the bend substrate provides tensile strain. The Radius of Curvature $\rho$ is calculated using the formula $(\rho - t)^2 = \rho^2 + (l/2)^2$ (c) Plot of strain vs number of turns. (d) Raman spectra for the monolayer MoS$_2$ sample for various levels of applied strain levels. Inset: illustration of the $E^1_{2g}$ and $A_{1g}$ vibrational modes. (e) Raman spectra of the bilayer MoS$_2$ sample for various strains. (f) & (g) Variation in the Raman shifts the $E^1_{2g}$ and the $A_{1g}$, for the monolayer and the bi-layer samples respectively as function of strain.

Strain has shown to have a profound effect on the bandgap of TMDC materials and, PL spectroscopy has been used monitor and characterize the strain induced bandgap variation in these systems[11,13,39]. In this session, we characterize the strain induced bandgap change on our mono- and the bilayer samples using PL spectroscopy. The results thus obtained are used to get more insights into and analyse the electrical transport data presented in the subsequent session. Fig. 2 (a) and (b) show representative PL spectra taken as a function of applied tensile strain on mono-and bi-layer $MoS_2$ samples respectively. Optical images of the corresponding mono- and bi-layer samples are given in the Supporting Information S2. The monolayer $MoS_2$ is characterized by a single PL emission peak at ~ 1.84 eV corresponding to the direct inter-band transition at the K – K point in the Brillouin zone[42]. In contrast, the bilayer $MoS_2$ flake exhibits two peaks; the A peak at ~1.84 eV corresponding to the direct transition at K – K point while the I-peak at ~ 1.56 eV is due to the indirect transition at the K - Γ point in the Brillouin zone[43]. We observe a redshift in the PL peaks of both the mono- and the bi-layer samples signalling a reduction in the bandgap with strain. The bandgap vs applied strain extracted from the PL spectra for the mono- and the bi-layer samples are shown in Fig. 2 (c) and (d) respectively. The PL peak for the monolayer shows a red-shift of ~ 78 ± 4 meV/% strain. For the bilayer, the A and the I peak show red-shifts of ~34 ± 3 meV/% strain and ~155 ± 11 meV/% strain respectively. Our observation of strain-induced reduction in the bandgap is consistent with the existing theoretical and experimental reports[11–13,39] confirming reduction in the $MoS_2$ bandgap and an effective transfer of strain on to the $MoS_2$ samples from the bending apparatus.

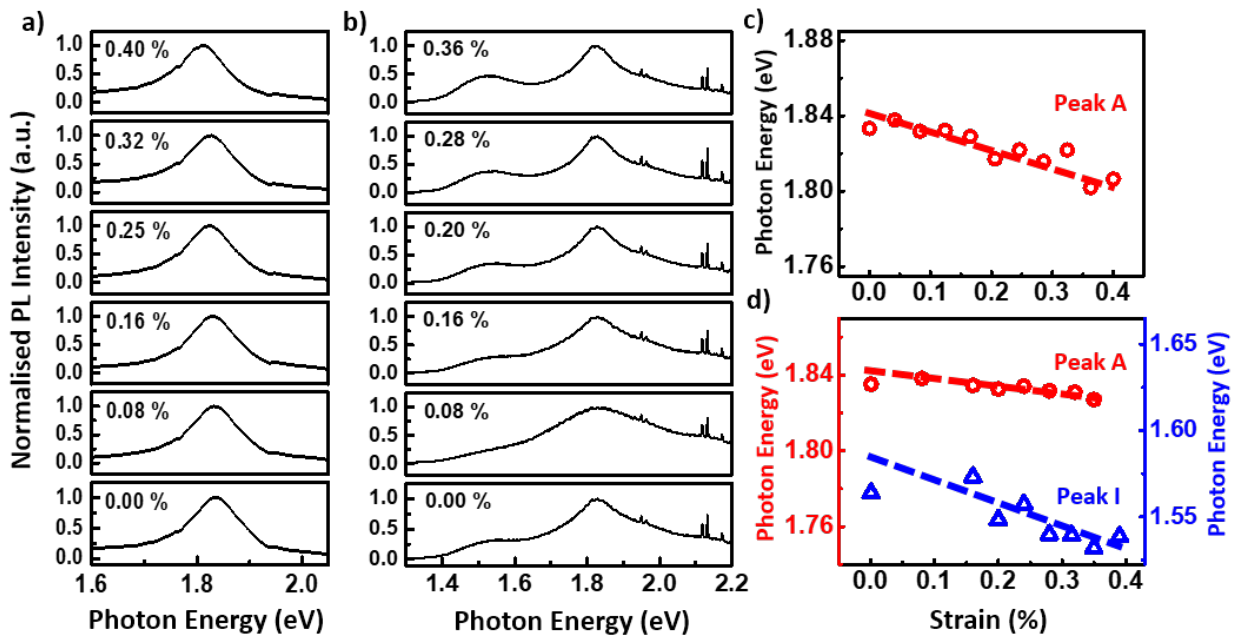

**Figure 2.** PL emission vs strain. (a) Evolution of PL spectra for monolayer $MoS_2$ sample showing the A peak - the direct transition at 1.84 eV for various values of applied strain (b) Evolution of PL spectra for the bi-layer $MoS_2$ sample showing the A peak - the direct transition at 1.84 eV and the I peak – the indirect transition at 1.56 eV for various values of applied strain (c) Variation in the A-peak position with strain (78meV/ % strain) extracted from Fig. 2 (a). (d) Variation in the A-peak (34meV/ % strain) and the I-peak (155meV/ % strain) positions with strain for the bi-layer $MoS_2$ device extracted from Fig. 2(b).

Any variation in the electrical properties such as bandgap, effective mass and carrier concentration need to reflect in the electrical transport characteristics of the material. Here, we investigate the influence of strain on the electrical transport characteristics using a bi-layer $MoS_2$ sample. Transport characterization conducted on a few-layered sample exhibiting similar behaviour is shown in the Supporting Information S3. Our monolayer $MoS_2$ samples on the polyimide substrate showed insulating behaviour. Surface roughness of substrate are shown to play a crucial role in reducing the conductivity of $MoS_2$[44,45] and, we believe that the substantial RMS surface roughness of the substrate, ~ 2 - 4 nm [see supporting information S4] made the samples nonconducting. Transport characterization conducted on a few-layered sample exhibiting similar behaviour is shown in the Supporting Information S3. Fig. 3 (a), open circles, show I-V characteristics of the bi-layer $MoS_2$ device for various percentage of applied strain. The inset shows the optical image of the device. The I-V characteristics show an enhancement in the current through the device with strain. The resistance of the device extracted from the linear low-bias regime of the I-V characteristics, shown in red-circles in Fig. 3 (b), shows a substantial reduction with strain. For the strain levels applied in this experiment, the bandgap change induced resistance variation dominates and the contribution from other effects such as the variation in the effective-mass are negligible[17,46]. For a semiconductor, the bandgap variation has an exponential influence on the carrier concentration and in-turn the resistance[47,48]. We estimate the change in the bandgap with strain, $\frac{\partial E_g}{\partial \epsilon}$, from the variation of resistance with strain, shown in Fig. 3 (b) red-circles, using the relation[48] $R = R_0 * exp\left[-\frac{\partial E_g}{\partial \epsilon}\frac{\epsilon}{2k_B T}\right]$. The $\frac{\partial E_g}{\partial \epsilon}$ ~ 195 m eV/%-strain thus obtained is considerably higher than ~ 155 m eV/%-strain, that we extracted from the PL spectroscopy for our bi-layer sample, and also exceeds those reported elsewhere[11,18]. Unlike the PL spectroscopy, the measurements discussed here probes not only the material but also the behaviour of the electrical contacts.

Metal-semiconductor contacts seldom yield Ohmic behaviour and most of the commonly used metals yield Schottky contacts with $MoS_2$[49]. The nature and behaviour of the contacts depend on the metal work-function, electron affinity of the semiconductor and also on the nature and density of the interface states[31].

The I-V characteristics shown in Fig. 3 (a) have two regimes; (i) a nearly linear low-bias regime and, (ii) a current-saturation regime at higher source-drain bias. This is atypical of a device with ohmic contacts and, the presence of a saturation region suggests the formation of potential barriers at the source and the drain contacts[50,51]. The saturation current has an exponential dependence on the barrier[31,47]. The rise in both the low-bias device conductance and the saturation current suggest that not only the device conductance but also the potential barriers formed at the electrical contact regions are modified by the strain.

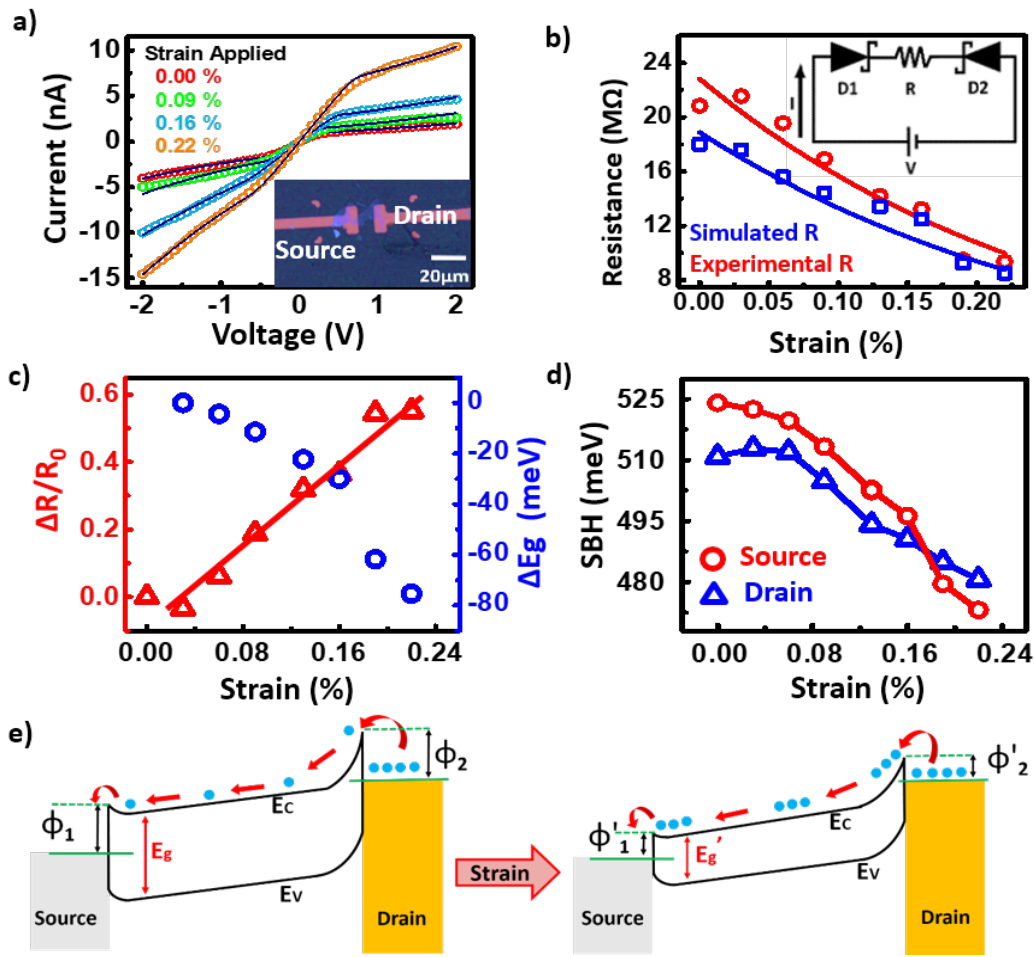

**Figure 3.** I-V characteristics. (a) Representative I-V characteristics for the bilayer $MoS_2$ device vs strain. Inset - optical image of the device. (b) Red open-circles show resistance vs strain extracted from the low-bias regime of the I-V characteristics and red-line shows an exponential fit giving $\frac{\partial Eg}{\partial \epsilon}$ ~195 m eV/% strain. The inset shows the circuit model used for the device simulation. Blue open-squares represent the resistance extracted from the simulation for different strains. (c) Change in $\frac{\Delta R}{R_0}$ (red open-triangles) and the bandgap $\Delta E_g$ (blue open-circles) with strain. (d) Change in SBH the source and the drain contacts, $\varphi_i$, with strain. (e) Band-diagram illustrating the prominent effects of strain on transport in the device. Strain causes a reduction in the bandgap resulting in (i) more carriers and reduction in the material resistance and (ii) the SBH – facilitating an enhancement in the transport.

To segregate the behaviour of the contacts and the material against strain, we analyse the data with the circuit model[22,50,52] shown in the inset to Fig. 3 (b). The model consists two Schottky diodes connected back-to-back representing the electrical contacts while the MoS$_2$ flake is modelled as a variable resistor connected between the diodes. Both the transport through the diodes and the resistor are influenced by the applied strain.

The I-V characteristics of the Schottky diode representing contacts is given by[47]

$$I_i(V_i) = I_{0,i} \, exp\left(\frac{qV_i}{\eta_i k_B T}\right)\left[1 - exp\left(\frac{qV_i}{k_B T}\right)\right]$$

where, $I_{0,i}$ is the reverse saturation current, $A_c$ is the contact area of the metal – semiconductor junction, $A^*$ is the Richardson's constant, $V_i$ is the voltage drop across each junction, $\eta_i$ is the ideality factor, $q$ is the electronic charge, $k_B$ the Boltzmann constant and, T is the temperature. The area of contact for the electrodes, ~ 400 μm$^2$, are determined from the optical image of the device. The initial guess to the resistance of the MoS$_2$ flake for each strain level is obtained from the linear part of the low-bias I-V characteristics. The initial guess to the $I_{0,i}$ are obtained from the current at the saturation region of the I-V characteristics. From the fit to the I-V traces corresponding to different strain values we extract the reverse saturation current and the ideality factor and the resistance of the MoS$_2$ flake. From the reverse saturation curve we extract the refined SBH using the relation[22,47], $\phi_i = \left(\frac{k_B T}{q}\right) \ln\left(\frac{A^* A_c T^2}{I_{oi}}\right)$. More details on the fitting procedure is given in the Supporting Information S5.

Cr/Au contacts on MoS$_2$ has been shown to exhibit Schottky behaviour and, the carrier injection across the barrier is governed by thermionic emission mechanism at room temperature[29,47,50]. Considering this, we have restricted the ideality factor $\eta_i$ of the junctions in the vicinity of 1. The goodness-of-fit is assessed by inspecting the coefficient-of-determination $R^2$ vs the $\eta_i$ and reverse saturation current $I_{0,i}$ as shown in the Supporting Information S6 and, we obtain a refined $\eta_i$ of 1.04, 1.01 for the source and the drain barriers.

The fits we obtained, shown in solid lines in Fig. 3(a), exhibit excellent agreement with the experimental I-V characteristics. The resistance of the MoS$_2$ as a function of strain, extracted from the fit is shown in the blue-squares in Fig. 3 (b). Assuming an exponential dependence of resistance to the strain[18,48] we obtain $\frac{\partial E_g}{\partial \epsilon} \sim 147$ meV/%-strain for the bi-layer sample, agreeing well with the PL data shown in Fig. 2 (d) and also concurs with other spectroscopic studies[11]. The variation in the device bandgap with strain extracted from the simulated resistance values is shown in blue open-circles and, the change in device resistance $\frac{\Delta R}{R_0}$ vs strain is shown in red open-triangles in Fig. 3(c), where $R_0$ is the resistance of the unstrained sample and $\Delta R$ is the change in resistance. From the resistance change we extract a piezoresistive gauge factor GF = $\frac{\Delta R}{R_0}/\epsilon \sim 321$ for the bi-layer device where $\epsilon$ is the strain on the sample. The GF for our bi-layer sample is substantially larger than that reported for Silicon[35].

Fig. 3 (d) shows the reduction in the SBH $\varphi_i$ obtained from the fit for the source and the drain barriers. The bilayer sample shows a reduction of ~ 55 meV/% strain in SBH. The few-layer sample shown in Supporting Information S3 shows a reduction of 40 meV/% strain. We infer that the major contribution to the SBH is the strain-induced bandgap reduction. The tensile strain results in lowering the conduction band minima at the Γ-point of the Brillouin zone, which will result in the reduction of the effective Schottky barrier at the source and the drain contacts[11,34]. Fig. 3 (e) illustrates the two dominant mechanisms contributing to the conductance modification (i) the bandgap induced change in the material resistance via carrier concentration enhancement (ii) the SBH reduction at the contacts, both will result in the enhancement of device conductance. Apart from the bandgap-change induced reduction in SBH, the contact-metal layer can also make better hybridization with the MoS$_2$ orbitals as a result of the reduction in the interlayer distance and weakening of the d$z^2$ orbitals[34]. The actual SBH in MoS$_2$-metal contacts are reported to be different from those predicted by the Schottky-Mott rule[31] and is mostly decided by Fermi-level pinning mechanisms[30,32]. In this scenario the ability to tune the SBH continuously by strain and engineer high efficiency electrical contacts will provide a major boost to the device engineering on MoS$_2$.

**Conclusions**

Strain can play a lead role in tuning the electrical properties of MoS$_2$ and other 2D materials for device engineering. There are only a few studies on probing the transport properties under the action of uniform strain across the sample. This manuscript is devoted to understand the effect of uniaxial in-plane tensile-strain on the electrical properties of MoS$_2$ samples. A three-point bending apparatus is used to induce uniform strain across the sample. Raman spectroscopy is used to infer and calibrate the strain on the lattice. The in-plane Raman modes showed a red-shift signaling softening of the in-plane phonon modes while the out-of-plane Raman modes remained relatively unaffected by the strain. A red-shift in the PL emission peaks corresponding to both the direct and indirect transitions is observed signaling a reduction in the bandgap with strain. Electrical transport studies with two-probe geometry are conducted on bi- and few-layered samples. Modulation of electrical conductivity and SBH at the electrical contacts against strain are the main themes of the manuscript. We observed a strong enhancement in the device conductance with strain for both mono and few-layered devices. Our bi-layer device exhibited a strong piezoresitive a GF ~ 321 superior to that of Silicon[35] while the few-layered device exhibited a GF ~ 48. In addition to the reduction in the flake resistance, we find a reduction in the source and the drain Schottky barriers heights. The tensile strain results in a continuous lowering of the conduction-band minima at the Γ-point of the Brillouin zone inducing a corresponding lowering of the Schottky barrier height[11,34]. Major contribution to the MoS$_2$ density of states at Fermi-level arise from the *d*-orbital of Mo atoms[34]. The increased overlap of the metal atom orbitals as a result of strain with these d-orbitals can also play a role in the reduction of the contact barrier heights. Resistance change against strain has a substantial contribution from the Schottky barrier modulation also. This effect can be used for engineering high-efficient, nanoscale strain sensors utilizing this effect. 2D systems offer a versatile platform to apply substantial amount of strain in a reversible manner. Mobility and conductivity enhancements, bandgap and SBH reductions with strain can bring a new perspective in device engineering on MoS$_2$, making it a potential platform for future microelectronics and straintronics technologies.

**Acknowledgements**

Authors acknowledge IISER Thiruvananthapuram for the facilities & infrastructure and, Rajeev N. Kini for the discussions and suggestions. MT acknowledges the funding received from the Science and Engineering Research Board, India under CRG12018/00421.

**Author Contributions**

MT conceived the problem. APJ and AT prepared and measured the devices and analyzed the data. MT and APJ prepared the manuscript.

**References**

1. Ieong, M., Doris, B., Kedzierski, J., Rim, K. & Yang, M. Silicon device scaling to the sub-10-nm regime. *Science* **306**, 2057-2060 (2004). doi:10.1126/science.1100731
2. Li, J., Shan, Z. & Ma, E. Elastic strain engineering for unprecedented materials properties. *MRS Bull.* **39**, 108–114 (2014).
3. Feng, J., Qian, X., Huang, C. W. & Li, J. Strain-engineered artificial atom as a broad-spectrum solar energy funnel. *Nat. Photonics* **6**, 866–872 (2012).
4. Hirth, J. P. & Lothe, J. *Theory of dislocations*. (Wiley, 1982).
5. Yue, Y., Liu, P., Zhang, Z., Han, X. & Ma, E. Approaching the theoretical elastic strain limit in copper nanowires. *Nano Lett.* **11**, 8,3151-3155 (2011). doi:10.1021/nl201233u
6. Brenner, S. S. Tensile strength of whiskers. *J. Appl. Phys.* **27**, 1484(1956). doi:10.1063/1.1722294
7. Lee, C., Wei, X., Kysar, J. W. & Hone, J. Measurement of the Elastic Properties and Intrinsic Strength of Monolayer Graphene. *Science* **321**, 385–388 (2008).
8. Bertolazzi, S., Brivio, J. & Kis, A. Stretching and Breaking of Ultrathin $MoS_2$. *ACS Nano* **5**, 9703–9709 (2011).
9. Cooper, R. C. *et al.* Nonlinear elastic behavior of two-dimensional molybdenum disulfide. *Phys. Rev. B* **87**, 035423 (2013).
10. Rice, C. *et al.* Raman-scattering measurements and first-principles calculations of strain-induced phonon shifts in monolayer $MoS_2$. *Phys. Rev. B* **87**, 081307 (2013).
11. Conley, H. J. *et al.* Bandgap Engineering of Strained Monolayer and Bilayer $MoS_2$. *Nano Lett.* **13**, 3626–3630 (2013).
12. Castellanos-Gomez, A. *et al.* Local strain engineering in atomically thin $MoS_2$. *Nano Lett.* **13**, 5361–5366 (2013).
13. He, K., Poole, C., Mak, K. F. & Shan, J. Experimental Demonstration of Continuous Electronic Structure Tuning via Strain in Atomically Thin $MoS_2$. *Nano Lett.* **13**, 2931–2936 (2013).
14. Desai, S. B. *et al.* Strain-Induced Indirect to Direct Bandgap Transition in Multilayer $WSe_2$. *Nano Lett.* **14**, 4592–4597 (2014).
15. Peelaers, H. & Van De Walle, C. G. Effects of strain on band structure and effective masses in $MoS_2$. *Phys. Rev. B - Condens. Matter Mater. Phys.* **86**, 1–5 (2012).
16. Johari, P. & Shenoy, V. B. Tuning the electronic properties of semiconducting transition metal dichalcogenides by applying mechanical strains. *ACS Nano* **6**, 5449–5456 (2012).
17. Ghorbani-Asl, M., Borini, S., Kuc, A. & Heine, T. Strain-dependent modulation of conductivity in single-layer transition-metal dichalcogenides. *Phys. Rev. B* **87**, 235434 (2013).
18. Manzeli, S., Allain, A., Ghadimi, A. & Kis, A. Piezoresistivity and Strain-induced Band Gap Tuning in Atomically Thin $MoS_2$. *Nano Lett.* **15**, 5330–5335 (2015).
19. Song, S. *et al.* Room Temperature Semiconductor-Metal Transition of $MoTe_2$ Thin Films Engineered by Strain. *Nano Lett. **16**, 188-193* (2016). doi:10.1021/acs.nanolett.5b03481
20. Yang, S. *et al.* Tuning the Optical, Magnetic, and Electrical Properties of $ReSe_2$ by Nanoscale Strain Engineering. *Nano Lett.* **15**, 1660–1666 (2015).
21. Chen, J.-H. *et al.* Piezoelectric effect in chemical vapour deposition-grown atomic-monolayer triangular molybdenum disulfide piezotronics. *Nat. Commun.* **6**, 1–8 (2015).
22. Quereda, J., Palacios, J. J., Agraït, N., Castellanos-Gomez, A. & Rubio-Bollinger, G. Strain engineering of Schottky barriers in single- and few-layer $MoS_2$ vertical devices.

*2D Mater.* **4**, 021006 (2017).
23. Wu, W. *et al.* Piezoelectricity of single-atomic-layer MoS$_2$ for energy conversion and piezotronics. *Nature* **514**, 470–474 (2014).
24. Shen, T., Penumatcha, A. V. & Appenzeller, J. Strain Engineering for Transition Metal Dichalcogenides Based Field Effect Transistors. *ACS Nano* **10**, 4712–4718 (2016).
25. Neri, I. & López-Suárez, M. Electronic transport modulation on suspended few-layer MoS$_2$ under strain. *Phys. Rev. B* **97**, 241408 (2018).
26. Zhang, Z. *et al.* Strain-Modulated Bandgap and Piezo-Resistive Effect in Black Phosphorus Field-Effect Transistors. *Nano Lett.* **17**, 6097–6103 (2017).
27. Liu, H., Neal, A. T. & Ye, P. D. Channel Length Scaling of MoS$_2$ MOSFETs. *ACS Nano* **6**, 8563–8569 (2012).
28. Das, S., Chen, H.-Y., Penumatcha, A. V. & Appenzeller, J. High Performance Multilayer MoS$_2$ Transistors with Scandium Contacts. *Nano Lett.* **13**, 100–105 (2013).
29. Moon, B. H. *et al.* Junction-Structure-Dependent Schottky Barrier Inhomogeneity and Device Ideality of Monolayer MoS$_2$ Field-Effect Transistors. *ACS Appl. Mater. Interfaces* **9**, 11240–11246 (2017).
30. Kim, C. *et al.* Fermi Level Pinning at Electrical Metal Contacts of Monolayer Molybdenum Dichalcogenides. *ACS Nano* **11**, 1588 - (2017). doi:10.1021/acsnano.6b07159
31. Rhoderick, E. H. & E. H. Rhoderick and Williams R.H. *Metal-semiconductor contacts.* (Clarendon Press, 1988).
32. Tung, R. T. Chemical Bonding and Fermi Level Pinning at Metal-Semiconductor Interfaces. *Phys. Rev. Lett.* **84**, 6078 (2000). doi:10.1103/PhysRevLett.84.6078
33. Chen, J. R. *et al.* Control of Schottky barriers in single layer MoS$_2$ transistors with ferromagnetic contacts. *Nano Lett.* **13**, 3106-3110 (2013). doi:10.1021/nl4010157
34. Liu, B., Wu, L.-J., Zhao, Y.-Q., Wang, L.-Z. & Cai, M.-Q. Tuning the Schottky barrier height of the Pd-MoS$_2$ contact by different strains. *Phys. Chem. Chem. Phys.* **17**, 27088–27093 (2015).
35. Kanda, Y. Piezoresistance effect of silicon. *Sensors Actuators A. Phys.* **9**, 2857-2861 (1991). doi:10.1016/0924-4247(91)85017-I
36. Castellanos-Gomez, A. *et al.* Deterministic transfer of two-dimensional materials by all-dry viscoelastic stamping. *2D Mater.* **1**, 11002 (2014).
37. Christopher, J. W. *et al.* Monolayer MoS$_2$ Strained to 1.3% with a microelectromechanical system. *J. Microelectromechanical Syst.* **28**, 2, 254 (2019). doi:10.1109/JMEMS.2018.2877983
38. Wang, Y., Cong, C., Qiu, C. & Yu, T. Raman spectroscopy study of lattice vibration and crystallographic orientation of monolayer MoS$_2$ under uniaxial strain. *Small* **9**, 2857-2861 (2013). doi:10.1002/smll.201202876
39. Zhu, C. R. *et al.* Strain tuning of optical emission energy and polarization in monolayer and bilayer MoS$_2$. *Phys. Rev. B* **88**, 121301 (2013).
40. Lee, C. *et al.* Anomalous Lattice Vibrations of Single- and Few-Layer MoS$_2$. *ACS Nano* **4**, 2695–2700 (2010).
41. Abin Varghese, C. H. S. M. T. Topography preserved microwave plasma etching for top-down layer engineering in MoS$_2$ and other van der Waals materials. *Nanoscale* **9**, 1–8 (2017).
42. Splendiani, A. *et al.* Emerging photoluminescence in monolayer MoS$_2$. *Nano Lett.* **10**, 1271-1275 (2010). doi:10.1021/nl903868w
43. Mak, K. F., Lee, C., Hone, J., Shan, J. & Heinz, T. F. Atomically Thin MoS$_2$: A New Direct-Gap Semiconductor. *Phys. Rev. Lett.* **105**, 136805 (2010).
44. Lee, G. H. *et al.* Flexible and transparent MoS$_2$ field-effect transistors on hexagonal

boron nitride-graphene heterostructures. *ACS Nano* (2013). doi:10.1021/nn402954e
45. Man, M. K. L. *et al.* Protecting the properties of monolayer $MoS_2$ on silicon based substrates with an atomically thin buffer. *Sci. Rep.* **6,** *20890* (2016). doi:10.1038/srep20890
46. Yu, S., Xiong, H. D., Eshun, K., Yuan, H. & Li, Q. Phase transition, effective mass and carrier mobility of $MoS_2$ monolayer under tensile strain. *Appl. Surf. Sci.* **325**, 27–32 (2015).
47. Sze, S. M. & Ng, K. K. *Physics of Semiconductor Devices: Third Edition. Physics of Semiconductor Devices: Third Edition* (Wiley India, 2009).
48. Vidana, A. *et al.* Conductivity modulation in strained transition-metal-dichalcogenides via micro-electro-mechanical actuation. *Semicond. Sci. Technol.* **34**, 045013 (2019).
49. Liu, Y., Stradins, P. & Wei, S.-H. H. Van der Waals metal-semiconductor junction: Weak Fermi level pinning enables effective tuning of Schottky barrier. *Sci. Adv.* **2**, 4 e1600069 (2016).
50. Chiquito, A. J. *et al.* Back-to-back Schottky diodes: The generalization of the diode theory in analysis and extraction of electrical parameters of nanodevices. *J. Phys. Condens. Matter* **24**, 225303 (2012). doi:10.1088/0953-8984/24/22/225303
51. Xu, W., Chin, A., Ye, L., Ning, C.-Z. & Yu, H. Electrical and optical characterization of individual GaSb nanowires. in *Quantum Dots, Particles, and Nanoclusters VI - Proceedings of SPIE- The international Society for Optical Engineering* **7224** (2009) 72240G. doi:10.1117/12.816931
52. Di Bartolomeo, A. *et al.* Asymmetric Schottky Contacts in Bilayer $MoS_2$ Field Effect Transistors. *Adv. Funct. Mater.* **28**, 1800657 (2018). doi:10.1002/adfm.201800657

# Supporting Information

**Strain-engineering the Schottky barrier and electrical transport on MoS₂**

Ashby Phillip John, Arya Thenapparambil and Madhu Thalakulam

School of Physics, Indian Institute of Science Education & Research Thiruvananthapuram
Kerala, India – 695551

## Table of contents:
1. Layer number determination of MoS₂ on polyimide substrate using Raman Spectra
2. Optical images of the flakes used for the mono and the bi-layered flakes
3. Transport characteristics of the few-layer sample
4. AFM Profile of Polyimide Substrate
5. Device Simulation
6. Goodness of the simulation

## 1. Layer number determination of MoS₂ on polyimide substrate using Raman Spectra

Raman spectra for mono-, bi- and few layered MoS₂ on the Polyimide substrate is given below. The spacing between $E_{2g}^1$ and $A_{1g}$ for mono (Red), bi (Blue) and Few (Green) layer on the flexible substrate is found to be 19.2 cm⁻¹, 22.03 cm⁻¹ and 25.41 cm⁻¹ respectively. Raman spectra for few layered MoS₂ on Si/SiO₂ (Violet) 25.17 cm⁻¹ is also shown for comparison. The Raman shifts observed on our samples are consistent with those on similar substrates reported elsewhere as tabulated below.

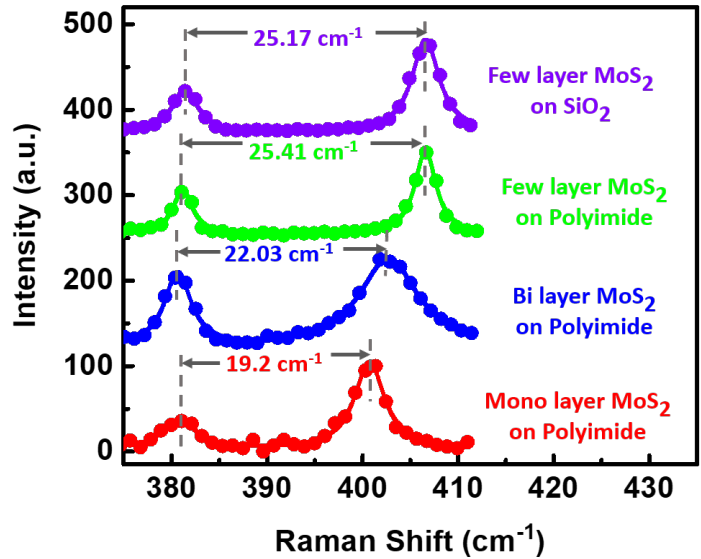

*Figure S1.* Raman Spectra for MoS₂ crystal of varying thickness transferred on to a Polyimide substrate. Spectra for Mono layer (Red), Bi layer (Blue) and Few layer (Green) MoS₂ has indicated, $E_{2g}^1$ and $A_{1g}$ spacing of 19.2 $cm^{-1}$, 22.03 $cm^{-1}$ and 25.41 $cm^{-1}$ respectively. Violet plot for Few Layer MoS₂ on Si/SiO₂ (25.1 $cm^{-1}$) is shown for reference.

**Table:** Reports on the usage of Raman spectra for layer number determination of $MoS_2$ on flexible substrates.

| Journal Title | Journal Ref | Layer Thickness | Substrate | Raman Shift | |
|---|---|---|---|---|---|
| Strain tuning of optical emission energy and polarization in monolayer and bilayer $MoS_2$ | C.R. Zhu et. al., PHYSICAL REVIEW B **88**, 121301(R) (2013) | 1L, 2L | PET | 1L | 18.7 cm$^{-1}$ |
| | | | | 2L | 22 cm$^{-1}$ |
| Bandgap Engineering of Strained Monolayer and Bilayer $MoS_2$ | Hiram J. Conley et.al., Nano Lett. 2013, **13**, 8, 3626-3630 | 1L 2L | SU8 deposited on Polycarbonate | 1L | 19 cm$^{-1}$ |
| | | | | 2L | 22 cm$^{-1}$ |
| Raman Spectroscopy Study of Lattice Vibration and Crystallographic Orientation of Monolayer $MoS_2$ under Uniaxial Strain | Y. Wang et. al., Small 2013, **9**, No. 17, 2857–2861 | 1L | PET | 1L | 19cm$^{-1}$ |

## 2. Optical images of the Mono and the Bi-layer $MoS_2$ flakes used for Raman and Photoluminescence measurement

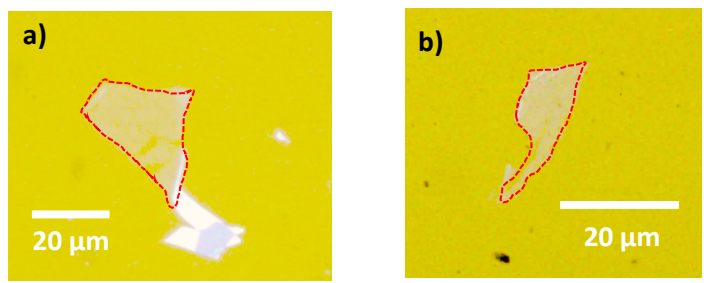

***Figure S2.*** *Optical image of $MoS_2$ crystal transferred on to a Polyimide substrate (a) mono layer (b) bi layer $MoS_2$ crystals used for Raman and PL measurement.*

## 3. Transport characteristics of the few-layer sample

Electrical transport measurement on a few layer $MoS_2$ device is given in Figure S3(a). Current through a device is governed by Schottky barrier and the resistance of the device. The change in Resistance is shown in Figure S3(b). The Red curve in Figure S3(b) indicates resistance extracted in the low bias regime of the I-V and the black curve is the resistance obtained from the simulation for various strain values. Figure S3(c) shows the change in band gap with strain. Schottky barrier height modification for Source and Drain junctions is shown in SI S3(d).

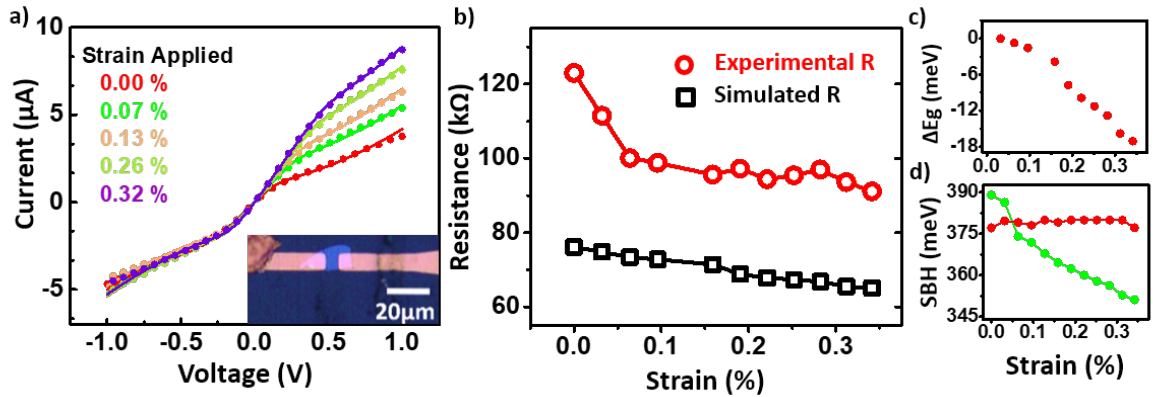

*Figure S3.* (a) I-V characteristics for a few-layer MoS$_2$ device. The solid circles are the experimental data and the lines are the simulated data. Optical image of the device is shown in the inset. (b) Piezoresistive response of the device. Red trace is the resistance extracted from the low-bias regime of the experimental data and black open-squares represent the resistance extracted from the simulated data. (c) Change in the Bandgap with strain obtained from the resistance change. (d) Change in SBH barrier for each junction with strain.

## 4. AFM Profile of Polyimide Substrate

Roughness of the Polyimide substrate used for the experiments are found to be around 2- 4nm. Height profiles obtained from different portions of the substrate is taken to estimate the RMS roughness of the substrate.

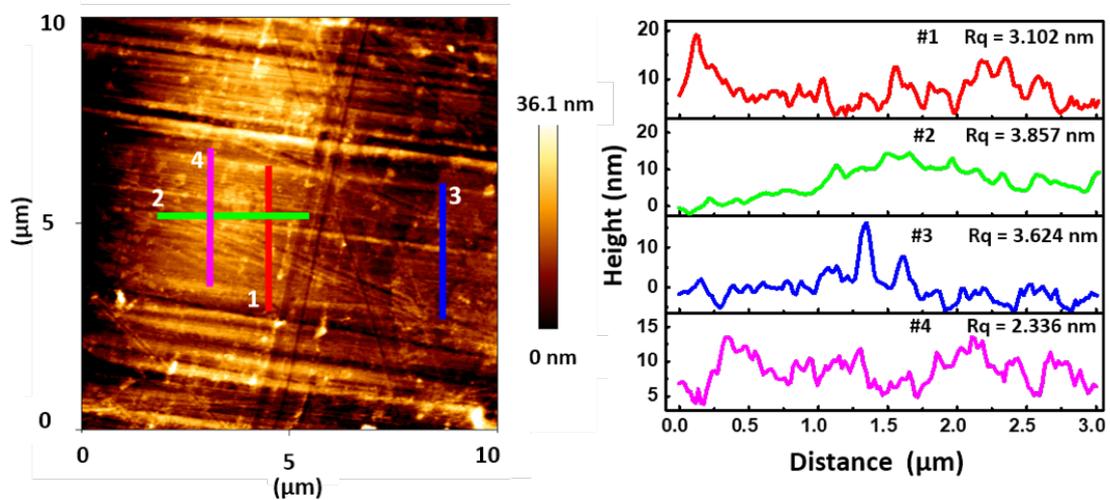

*Figure S4.* AFM surface profile for the flexible substrate. Right hand side shows the line cut taken from different regions of the substrate. The RMS surface roughness varies from 2 to 4nm.

## 5. Device Simulation

Schottky Barriers and the bulk resistance of a device can be modelled with back to back Schottky Diodes and a series resistor, R [1]. The circuit used for the simulation is given below

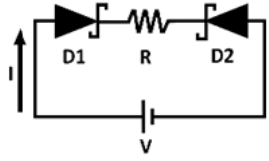

*Figure S4. Schematic representation of a back to back Schottky Diode with series resistor used for the modelling and Simulating the I-V curves.*

The Schottky Diode equation for each of the barrier is given by

$$I_i(V_i) = I_{0i} * exp\left(\frac{qV_i}{\eta_i k_B T}\right)\left[1 - exp\left(-\frac{qV_i}{\eta_i k_B T}\right)\right]$$

where the reverse saturation current is given by $I_{0i} = A_c A^* T^2 * exp\left(\frac{q\phi_i}{k_B T}\right)$

$V_i$ is the voltage drop across each Schottky barrier, $\eta_i$ is the ideality factor for the Schottky diode, $q$ is the electronic charge, $k_B$ Boltzmann constant and T is the temperature at which the measurement is taken - 300K.

$A_c$ is the contact area of the Metal – Semiconductor Junction. A* is the Richardson's constant and $\varphi_i$ is the Schottky barrier height.

According to the model, each Diode attributes the features of a metal semiconductor junction. The following parameters – $I_{01}$, $I_{02}$, $\eta_1$, $\eta_2$ and R is extracted from the fit where the subscript 1 and 2 are used to identify the junctions.

The total voltage is the sum of the voltage drop across each junction and the bulk resistance.

$$V = V_{D1} + V_{D2} + I_T(V) * R$$

The current through the device is determined by the voltage drop across each diode. When one diode is reversed biased and the other is forward biased and the current through the forward biased diode is controlled by the reversed biased diode.

$$I_{D1}(V_{D1}) = - I_{D2}(V_{D2}) = I_T$$

$I_T$ is the current flowing through the device which should be equal to the current through each junction. $V_{D1}$ and $V_{D2}$ is the voltage drop across junction 1 and 2 respectively.

$$I_{D1}(V_{D1}) = - I_{D2}(V - V_{D1} - I_{D1}(V_{D1}) * R)$$

Here we have two independent variables $V_{D1}$ and V to fit the equation.

$V_{D1}$ is found out by solving the above three equations for different parameters namely $I_{01}$, $I_{02}$, $\eta_1$, $\eta_2$ and R and put back into the equation to fit for the experimental data. This process is repeated until we get a good fit.

## 6. Goodness of the simulation

Coefficient-of-determination $R^2$ is calculated using the following formula[2]

$$R^2 = 1 - \frac{\sum(Observed - Predicted)^2}{\sum(Observed - Mean\ Observed)^2}$$

$R^2$ gives an impression how varied the simulated data are from the actual data and how good the parameter set used for the simulation is close to the actual parameters extracted from the experimental data. A 2D colour plot for $R^2$ with $\eta_i$ and $I_{0i}$ on the x and y axis respectively for the bilayer device described in the main article is given below. Figure S5(a) and S5(b) are the colour plot of $R^2$ for the metal-semiconductor junctions on the source and drain regions respectively for 0.00 % strain. Similarly figure S5(c) and S5(d) is for 0.28 % strain. The white dot in each figure shows the values of the refined $\eta_i$ and $I_{0i}$ used for the simulation.

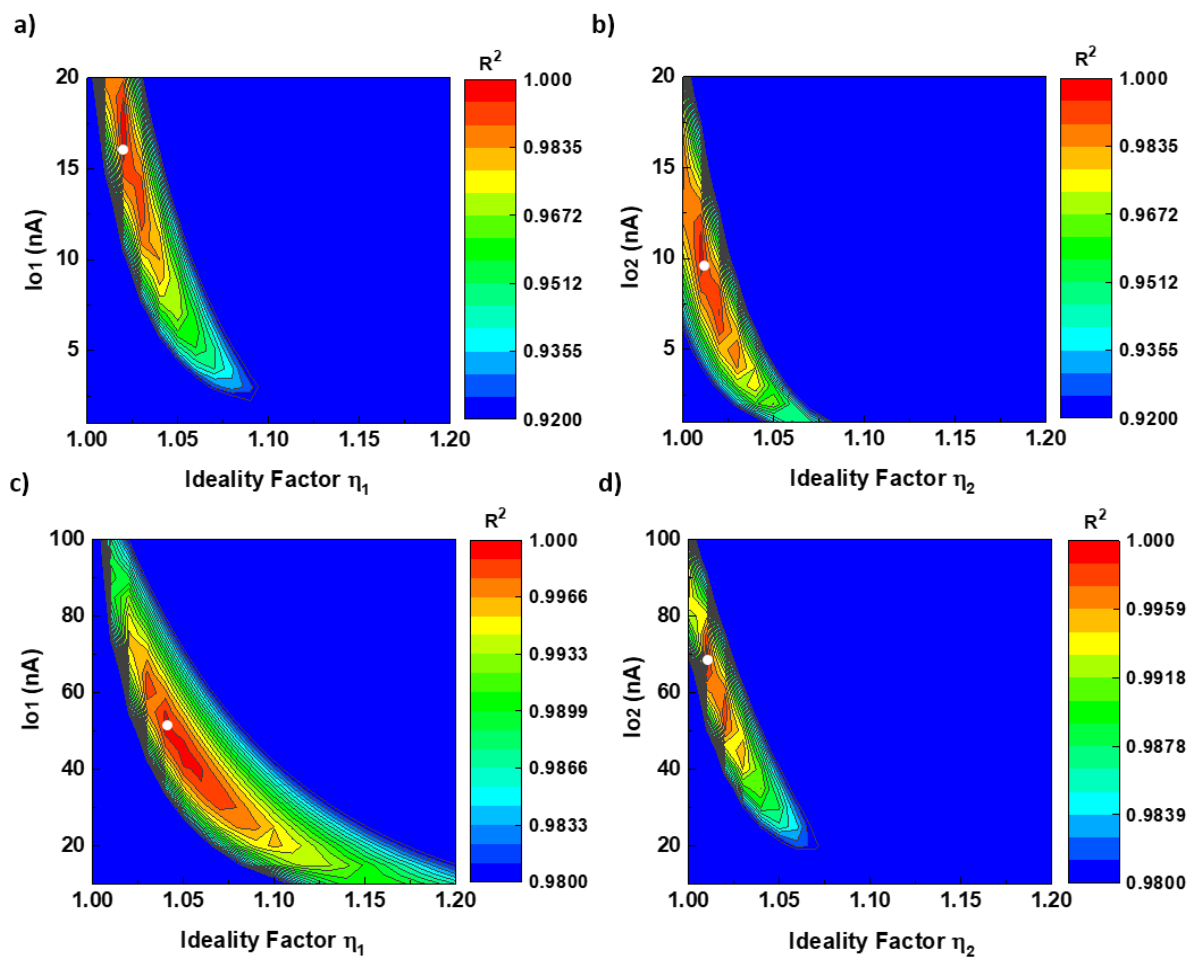

***Figure S5.*** *(a) and (b) represents 2D colour plot for the $R^2$ vs $\eta_1$ and $I_{o1}$ for the I-V with 0.00 strain % for the source and the drain metal – semiconductor junctions. (c) and (d) represent $R^2$ vs $I_{o2}$ and $\eta_2$ for 0.28 % strain for the same junctions. The white dot on the panels indicate the $I_{o2}$ and $\eta_2$ used for the simulation to get the best fit to the data.*

# Reference

1   Quereda, J., Palacios, J. J., Agräit, N., Castellanos-Gomez, A. & Rubio-Bollinger, G. Strain engineering of Schottky barriers in single- and few-layer MoS$_2$ vertical devices. *2D Mater.* **4**, (2017)
2   A. Colin Cameron and F. A. G. Windmeijer, J. Econom. **77**, 329 (1997).